\documentclass[a4paper,12pt]{article}
\usepackage[utf8]{inputenc}
\usepackage{graphicx}
\usepackage[labelfont=sc,font=footnotesize,margin=18pt]{caption}
\usepackage{physics}
\usepackage{amssymb}
\usepackage{bbm}
\usepackage[amssymb]{SIunits}
\usepackage{cite}
\usepackage[margin=1in]{geometry}
\usepackage{setspace}
\renewcommand{\vec}[1]{\mathbf{#1}}
\newcommand{\bleq}{\ensuremath{\mathrel{\phantom{=}}}}
\newcommand{\nnl}{\nonumber\\}

\newcommand{\rmi}{\mathrm{i}}
\newcommand{\rme}{\mathrm{e}}

\title{Gravitational entanglement and the mass contribution of internal energy in nonrelativistic quantum systems}
\author{Andr\'e Gro{\ss}ardt}
\date{\parbox{\linewidth}{\small\centering%
Friedrich Schiller University Jena\\
Institute for Theoretical Physics \\
Fröbelstieg 1, 07743 Jena, Germany\\
andre.grossardt@uni-jena.de
\endgraf\bigskip
Essay written for the Gravity Research Foundation\\2022 Awards for Essays on Gravitation.
}}

\begin{document}
\maketitle
\begin{abstract}
Recently, interest has increased in the entanglement of remote quantum particles through the Newtonian gravitational interaction, both from a fundamental perspective and as a test case for the quantization of gravity. Likewise, post-Newtonian gravitational effects in composite nonrelativistic quantum systems have been discussed, where the internal energy contributes to the mass, promoting the mass to a Hilbert space operator. Employing a modified version of a previously considered thought experiment, it can be shown that both concepts, when combined, result in inconsistencies, reinforcing the arguments for the necessity of a rigorous derivation of the nonrelativistic limit of gravitating quantum matter from first principles.
\end{abstract}
\thispagestyle{empty}
\clearpage
\setcounter{page}{1}

\doublespacing

Rapid progress in quantum technologies has shifted the spotlight on proposals~\cite{carlipQuantumGravityNecessary2008,zychQuantumInterferometricVisibility2011,pikovskiProbingPlanckscalePhysics2012,yangMacroscopicQuantumMechanics2013,grossardtOptomechanicalTestSchrodingerNewton2016,belenchiaTestingQuantumGravity2016,boseSpinEntanglementWitness2017,marlettoGravitationallyInducedEntanglement2017,westphalMeasurementGravitationalCoupling2020,howlNonGaussianitySignatureQuantum2021} for laboratory experiments that, through various approaches, observe aspects of the gravitational interaction between quantum systems in nonclassical states. The hope of their proponents is, that such experiments could provide new insights into the interplay of gravitational and quantum physics that may pave the way for a complete and fully consistent theory; or at least confirm some expected, yet untested principles for quantum gravity. Although the incompleteness of classical general relativity in conjunction with quantum field theory is evident at the Planck scale, it may appear that these low-energy situations pose no problem to the theories and methods currently available in fundamental physics. Quantum field theory in curved spacetime~\cite{birrellQuantumFieldsCurved1982,waldQuantumFieldTheory1994,barQuantumFieldTheory2009,brunettiAdvancesAlgebraicQuantum2015} provides a consistent framework for quantum matter in external gravitational fields, that is, as long as the backreaction of matter onto spacetime can be neglected. The perturbative quantization of general relativity~\cite{feynmanQuantumTheoryGravitation1963,donoghueGeneralRelativityEffective1994,burgessQuantumGravityEveryday2004} as an effective low-energy theory can explain the gravitational interaction between quantum systems in close analogy to the electromagnetic forces. One merely needs to turn a blind eye to its nonrenormalizable divergences at high energies.

The theoretical description of laboratory experiments, ordinarily, makes use of neither of these frameworks. Instead, these experiments are modeled by solutions of a nonrelativistic Schrödinger equation of some kind, with the gravitational interaction being introduced in a more or less ad-hoc manner. The correct Schrödinger equation should, of course, be derivable from the proper relativistic theory, be it quantum fields in curved spacetime or perturbative quantum gravity. It turns out, however, that this derivation is anything but straightforward, and of the obstacles one encounters along the way to the right description of laboratory quantum systems in weak, nearly-Newtonian gravitational potentials two, in particular, stand out.

First, there is no unambiguous answer to the question of how quantum matter sources the gravitational field. Despite the possibility of a perturbative quantum field theory at low energies, in the absence of a conclusive theoretical argument for the necessity of quantization~\cite{kibbleSemiClassicalTheoryGravity1981,huggettWhyQuantizeGravity2001,mattinglyQuantumGravityNecessary2005,albersMeasurementAnalysisQuantum2008,kentSimpleRefutationEppley2018,grossardtThreeLittleParadoxes2022}, mayhaps the gravitational field of atoms or molecules favors being modeled by the semiclassical point of view~\cite{mollerTheoriesRelativistesGravitation1962,rosenfeldQuantizationFields1963}, where a single, classical spacetime is coupled to quantum matter fields. The absence of evidence for quantization also triggered the development of other alternatives to a quantized theory~\cite{albersMeasurementAnalysisQuantum2008,kafriClassicalChannelModel2014,tilloySourcingSemiclassicalGravity2016,oppenheimPostquantumTheoryClassical2021}.
Raising the question: which types of experiments might confirm the quantum character of gravity? A signature feature could be the capability of the gravitational interaction to entangle distant quantum states~\cite{kafriClassicalChannelModel2014,boseSpinEntanglementWitness2017,marlettoGravitationallyInducedEntanglement2017}. Despite some controversy~\cite{hallTwoRecentProposals2018,marconatoVindicationEntanglementbasedWitnesses2021,hallCommentVindicationEntanglementbased2021,marlettoResponseCommentVindication2021,anastopoulosGravitationalEffectsMacroscopic2021} about which alternative models would be falsified by witnessing gravitationally induced entanglement, there seems to be a broad consensus that perturbative quantum gravity does, in fact, predict such entanglement~\cite{danielsonGravitationallyMediatedEntanglement2021,christodoulouLocallyMediatedEntanglement2022}.

Second, when it comes to the description of quantum systems in a curved background spacetime to post-Newtonian order, i.\,e. relativistic corrections to the Schrödinger equation, there has been some discussion~\cite{zychQuantumInterferometricVisibility2011,zychGeneralRelativisticEffects2012,pikovskiUniversalDecoherenceDue2015,bonderQuestioningUniversalDecoherence2016,bonderCanGravityAccount2015,pangUniversalDecoherenceGravity2016,diosiCentreMassDecoherence2017,pikovskiTimeDilationQuantum2017,schwartzPostNewtonianHamiltonianDescription2019} as to properly modeling systems with internal degrees of freedom. Although quantum field theory on curved spacetime should be applicable in these situations, no practicable derivation of observable effects has been established. 
General relativity teaches us that all energy gravitates, and it has been suggested that a quantum system with internal energy couples to a Newtonian potential with a passive gravitational mass\footnote{Planck units with $c = G = \hbar = 1$ are used throughout this essay.} \emph{operator} $\hat{M} = m \,\hat{\mathbbm{1}} + \hat{H}_\text{int}$, acting on the internal Hilbert space $\mathcal{H}_\text{int}$. Based on this assumed coupling, multiple interesting effects have been discussed for the case that the internal state becomes entangled with the system's center of mass, including dephasing~\cite{zychQuantumInterferometricVisibility2011,zychGeneralRelativisticEffects2012} and decoherence~\cite{pikovskiUniversalDecoherenceDue2015,pikovskiTimeDilationQuantum2017} due to gravitational time dilation, and even a reformulation of the equivalence principle for quantum matter~\cite{zychQuantumFormulationEinstein2018}. 

Let us assume a system in an eigenstate $\ket{\Psi} = \ket{\psi} \otimes \ket{E}$ of the internal energy with mass eigenvalue $M = m + E$. Then being exposed to the Newtonian potential $\Phi$, we see that the center-of-mass state $\ket{\psi} \in \mathcal{H}_\text{cm}$ obeys the Schrödinger equation
\begin{equation}\label{eqn:se-grav}
 \rmi \partial_t \ket{\psi} = \left( \frac{\hat{\vec p}^2}{2M} + M \Phi \right) \ket{\psi} \,.
\end{equation}
In fact, equation~\eqref{eqn:se-grav} is more than a mere suggestion; it has been experimentally confirmed for neutrons~\cite{colellaObservationGravitationallyInduced1975, nesvizhevskyMeasurementQuantumStates2003} as well as atoms~\cite{fixlerAtomInterferometerMeasurement2007}, for both of which the passive gravitational mass mainly consists of internal energy.
Whereas $M$ was a constant throughout these experiments, the treatment of mass as an operator leads one to consider it a dynamical degree of freedom that can be altered in the course of the experiment. Instead of the Galilean symmetry of the Schrödinger equation in nonrelativistic quantum mechanics, one finds that such a theory with a dynamical mass requires the central extension of the Galilei group with central element $M$ as a symmetry~\cite{bargmannUnitaryRayRepresentations1954,giuliniGalileiInvarianceQuantum1996}. There is no a-priori reason to question such dynamics, as long as one is comfortable with this change of the spacetime symmetry group.

Established principles encourage us to regard both the coupling of internal energy to the Newtonian potential and gravitationally induced entanglement as reasonable assumptions. Curiously, if taken together, these two assumptions can result in situations that allow for faster-than-light signalling. Can they be made compatible with each other?

\begin{figure}
 \centering
 \includegraphics[scale=0.8]{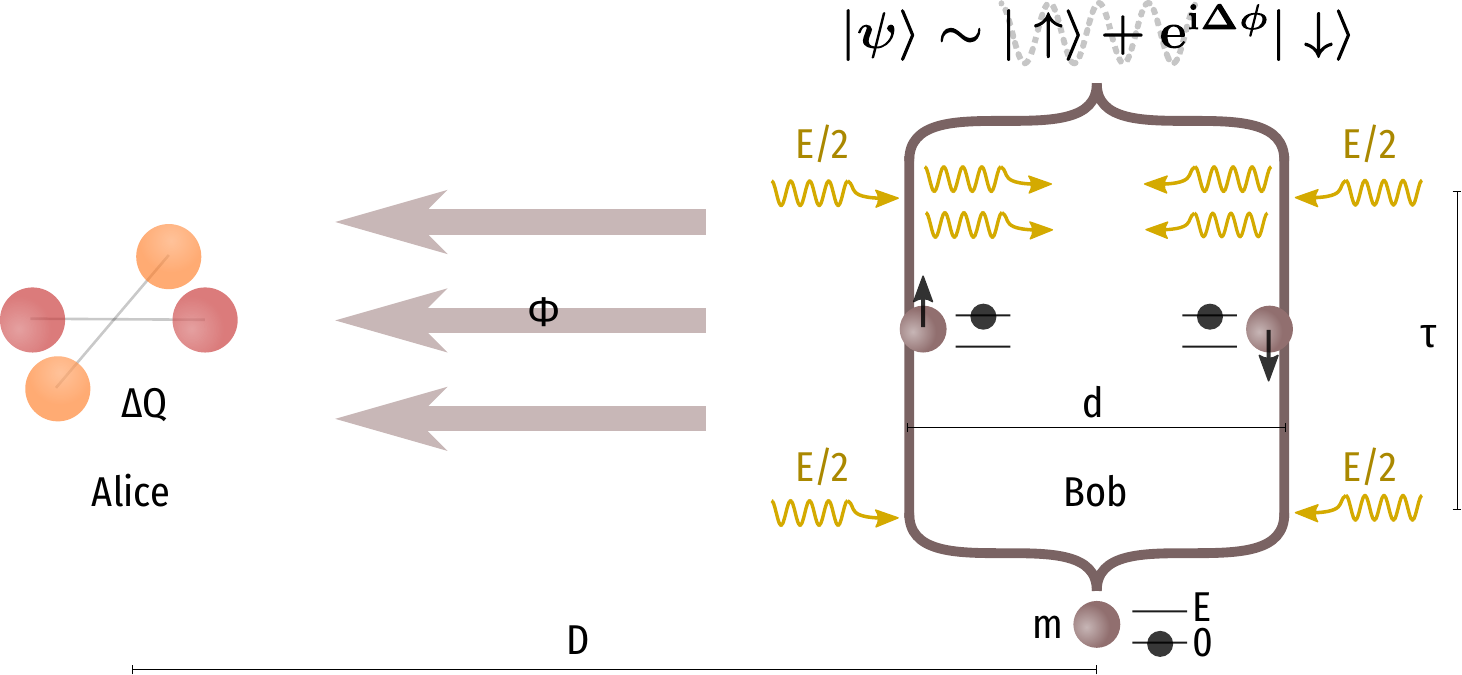}
 \caption{Schematics of the proposed thought experiment: Alice creates a superposition of massive particles resulting in a difference $\Delta Q$ in the gravitational quadrupole moment with respect to her center of mass. Bob has a massive particle of spin $\tfrac12$ and with an additional internal degree of freedom with two energy levels separated by $E$. The particle passes a Stern--Gerlach interferometer and absorbs two photons of total energy $E$ and zero total momentum. After the free flight time $\tau$ stimulated emission resets the internal degree of freedom to the ground state and the particle trajectories are recombined. A projective spin measurement finds a gravitational phase difference $\Delta \phi \sim E \Phi$.}
 \label{figure1}
\end{figure}

In order to uncover the conflict between gravitational entanglement and the gravitation of internal energy, we turn to a modified version of a previously examined thought experiment~\cite{mariExperimentsTestingMacroscopic2016,belenchiaQuantumSuperpositionMassive2018}. Consider the situation depicted in figure~\ref{figure1}, where Alice possesses a superposition state $\ket{\chi} \sim \ket{\chi_1} + \ket{\chi_2} \in \mathcal{H}_A$ of two mass distributions with a difference $\Delta Q = Q_1 - Q_2$ in their gravitational quadrupole\footnote{Conservation of the center of mass implies that the quadrupole moment is the lowest order nontrivial contribution to the gravitational potential for an observer outside of Alice's lab~\cite{belenchiaQuantumSuperpositionMassive2018}. If for symmetry reasons the lowest order contribution comes from higher moments~\cite{rydvingGedankenExperimentsCompel2021}, the arguments made in this essay apply accordingly.} moment. Bob, situated at a distance $D$, uses an interferometric experiment to sense the gravitational potential $\Phi_{1,2} \approx \Phi_0 -Q_{1,2}/r^3$ of Alice's quadrupole at distance $r$ from her center of mass.\footnote{The Newtonian potential at Bob's location will be retarded by a time $D$~\cite{christodoulouLocallyMediatedEntanglement2022}. This is irrelevant for the further considerations, because Alice can prepare her superposition an arbitrarily long time before the actual experiment begins. The retardation $d$ between Bob's trajectories is negligible and can be entirely mitigated by adjusting the trajectories.} Specifically, let Bob's system be a spin-$\tfrac12$ particle with the center-of-mass state $\ket{\psi} \in \mathcal{H}_\text{cm}$, initial spin state $\ket{+} = (\ket{\uparrow} + \ket{\downarrow})/\sqrt{2} \in \mathcal{H}_\text{spin}$, and an internal state modeled by a two-level system with basis $\{\ket{0},\ket{E}\} \subset \mathcal{H}_\text{int}$. A magnetic field gradient $\partial_z B$ is applied for a short time $\tau_a$ with a sign flip after half the time, resulting in parallel spatial trajectories at a distance $d \sim \mu \, \tau_a^2 \, \partial_z B / m$ entangled with the spin, where $\mu$ is the magnetic moment of the particle. The same procedure with inverted sign reunites the trajectories after a free flight time $\tau$. For $\tau_a \ll \tau$, the gravitational interactions during the acceleration phase are negligible. The distance $d$ can be varied independently from the time $\tau_a$ by varying the field gradient or the mass.

At the beginning of the free flight phase, the initial state $\ket{\Psi_i}$ in the total Hilbert space $\mathcal{H} = \mathcal{H}_A \otimes \mathcal{H}_\text{spin} \otimes \mathcal{H}_\text{cm} \otimes \mathcal{H}_\text{int}$ is
\begin{equation}\label{eqn:psii}
 \ket{\Psi_i} \sim \ket{\chi} \otimes \left(
  \ket{\uparrow} \otimes \ket{D-\tfrac{d}{2}} + \ket{\downarrow} \otimes \ket{D+\tfrac{d}{2}} 
 \right) \otimes \ket{0} \,,
\end{equation}
where we write $\ket{x} \in \mathcal{H}_\text{cm}$ for states sharply peaked around position $x$. 
The interaction with the Newtonian potential $\Phi_{1,2}$ sourced by Alice's superposition of masses, followed by the reunification of the trajectories, results in the nonlocally entangled state
\begin{align}\label{eqn:psif}
 \ket{\Psi_f} &\sim \Big[ 
 \ket{\chi_1} \otimes \left(\ket{\uparrow} 
 + \rme^{\rmi \Gamma} \ket{\downarrow} \right)
 \nnl &\bleq + \rme^{2 \rmi \Gamma D/(3d)} 
 \ket{\chi_2} \otimes \left( \rme^{\rmi \Gamma}\ket{\uparrow} 
 + \ket{\downarrow} \right)
 \Big] \otimes \ket{D} \otimes \ket{0}
\end{align}
with $\Gamma = 3 m \, \tau\, \Delta Q\, d / (2 D^4)$ and to linear order in $d \ll D$. For $\Gamma  = \tfrac\pi2$ the spin states become orthogonal, and after a projective spin measurement Bob is able to predict Alice's mass distribution with certainty. The reduced density matrix for Alice's state becomes diagonal, implying a complete loss of coherence, which Alice and Bob agree to use in order to encode a zero. If, instead, Bob decides not to perform the experiment, Alice finds her state in perfectly coherent superposition, encoding the number one as per Alice and Bob's agreement. Bob can use this protocol to send a binary encoded message to Alice.

Studied by Belenchia et al.~\cite{belenchiaQuantumSuperpositionMassive2018}, the field character of the gravitational interaction avoids faster-than-light signalling in this scenario. In order to be able to observe interference, Alice must avoid sending which-way information in the form of gravitational waves, implying that she must recombine her superposition slow enough, in time $T^2 > \Delta Q$. If Bob monitors the position of a test mass for time $\tau$, he finds a separation $\delta x \sim \Delta Q \,\tau^2 / D^4$. In the case $D > T + \tau$, in which faster-than-light signalling would be possible, this amounts to a separation $\delta x < 1$ below the Planck length. Rydving et al.~\cite{rydvingGedankenExperimentsCompel2021}, who consider the interferometric detection of Alice's field by Bob, similar to our thought experiment here, notice that it is the requirement that interference fringes be at least a Planck length apart, which excludes the possibility to signal superluminally in this case. For the thought experiment described above and depicted in figure~\ref{figure1}, the relative phase between the states $\ket{\uparrow}$ and $\ket{\downarrow}$ oscillates with a spatial period
\begin{equation}\label{eqn:fringe}
 \delta_f \sim \frac{\tau_a}{m \, d} 
 \sim \frac{\tau_a \, \tau \, \Delta Q}{D^4}  \,,
\end{equation}
where we already used the condition $\Gamma = \tfrac\pi2$. With $\Delta Q < T^2$ and the condition for faster-than-light signalling that all times are smaller than $D$, one finds that $\delta_f < 1$ is below the Planck length. When averaged over a spatial area of at least the Planck scale, Bob cannot extract any information about Alice's state from his measurement, and any signalling between them is avoided as long as the Planck length presents the expected fundamental limitation on spatial resolution. Given the right parameters, the protocol established between Alice and Bob allows the sending of signals; nonetheless, the combination of the Planck length limit, gravitational radiation, and quantum mechanical complementarity prevents any obvious conflicts with causality.

As yet, we have not made use of the internal degree of freedom of our particle. If the particle is in the excited state $\ket{E}$ throughout the experiment, we find the same relations \eqref{eqn:psii} through \eqref{eqn:fringe}, simply substituting the mass $m$ by $M = E + m$. The mass being a dynamical parameter, however, provides us with the possibility to alter it during the course of the experiment. We assume that shortly after the initial acceleration phase the particle absorbs two counter-propagating photons with energy $E/2$ each, promoting the initial state to
\begin{equation}\label{eqn:psiiprime}
 \ket{\Psi_i} \to \ket{\Psi_i'} \sim \ket{\chi} \otimes \left(
  \ket{\uparrow} \otimes \ket{D-\tfrac{d}{2}} + \ket{\downarrow} \otimes \ket{D+\tfrac{d}{2}} 
 \right) \otimes \ket{E} \,.
\end{equation}
Before recombining the two paths, the particle reverts to the internal ground state by a stimulated two-photon emission. 
During the free flight time $\tau$, the Newtonian potential is then expected to couple to the total mass $M$, including the internal energy, rather than to the rest mass. The gravitational phase shift acquired is proportional to $M$ rather than $m$, and we find the same final state \eqref{eqn:psif}, only with the replacement $m \to M$, that is, with the phase $\Gamma \to \Gamma' = (M/m) \Gamma$. The fringe visibility \eqref{eqn:fringe}, on the other hand, is determined by the acceleration phases during which the particle is in the internal ground state. It still depends on the rest mass $m$. Accordingly, the condition for faster-than-light signalling only requires a spacing of $\delta_f < M/m$ between interference fringes. Assuming $E \gg m$, this value can considerably exceed the Planck length. The arguments protecting us from problems with causality for particles without internal degrees of freedom~\cite{belenchiaQuantumSuperpositionMassive2018,rydvingGedankenExperimentsCompel2021} crumble.

It is important to note that quantization of matter is necessary for both ends of the argument. Alice's mass must be in a superposition state, such that the gravitational field can become entangled with it. Bob's particle, too, must be in a quantum state. Although a classical test particle with internal energy $E$ would gravitate with the total mass $M$ as well, only the superposition of different position states and the phase difference acquired between them makes the increase in the passive gravitational mass usable for a precise determination of the gravitational potential. The gravitational field, in contrast, need only be able to generate entanglement. Beyond this property, there is no requirement of quantization. Notably, it is irrelevant for the argument whether decoherence of Alice's particle occurs due to emission of gravitons or of classical gravitational waves.

The condition for faster-than-light signalling with a fringe spacing of at least the Planck length can be rephrased as
\begin{equation}
 \frac{1}{M}
 < \frac{D^3}{M \, \tau \, \Delta Q}
 \sim \frac{d}{D} \sim
 \frac{\tau_a}{m \, D \, \delta_f}
 < \frac{1}{m} \,.
\end{equation}
It is then evident, that it cannot be satisfied in the case $M = m$ of constant mass. For $M \gg m$, the required distance $d$ can be made small by increasing the internal energy. The particle trajectories are well approximated by test particle geodesics, as long as their distance is sufficiently far above the Schwarzschild radius of the mass, $d \gg M$.

In order to ensure that the emitted photons do not carry which-way information, we assume (deviating from the schematic drawing in figure~\ref{figure1}) that the photon beam is perpendicular to the plane formed by the interferometric paths and spread over an area $A \gtrsim d^2$ centered between them, such that absorption and emission take place with equal probability for both spin states. The probability of a successful experiment, in which the atom gets excited and reverts to the ground state, is then of the order of $(\sigma/A)^2$, where $\sigma$ is the cross section for two-photon absorption and stimulated emission. Successful measurements can be post-selected by choosing $\tau > d$ and timing the arrival of the photons. For any given two-level system with a mass $m$, an energy gap $E \gg m$ above the Planck energy, and a cross section $\sigma$ for absorption and stimulated emission of two photons whose square root, measured in Planck lengths, is larger than the energy gap, measured in Planck energies, one can always choose parameters such that $\sigma \sim d^2$, for which the probability for success approaches unity. For instance, a hydrogen-like ``atom'' built from two sub-Planckian masses, charged such that $q^2 \sqrt{m} \sim 137$, would do the trick with a Bohr radius of $a_0 \sim 1/\sqrt{m}$.
Especially for a small ratio $d/D$, the probability can be further increased by running a large number of identical experiments in parallel. If only in a single one of them the particle absorbs and re-emits the photons, Alice should not be able to produce an interference pattern, and Bob can signal faster than light over a maximum distance $D \lesssim M\,d$.

Is there another argument preventing us from performing this experiment, at least in principle? The analogy with electrodynamics, which has proven helpful previously~\cite{baymTwoslitDiffractionHighly2009,belenchiaQuantumSuperpositionMassive2018}, is of restricted use; it seems infeasible to add a significant electric charge to a particle during the course of an interference experiment without decohering its state, chiefly because the force between equal charges is repulsive. Contrary to the case with constant mass, which can be compared to the electromagnetic version in almost complete analogy, the coupling to a variable internal energy presents us with a situation genuinely unique for the gravitational interaction.

One could object that adding internal energy much larger than the rest mass turns the particle into a relativistic system which cannot be appropriately described by the nonrelativistic Schrödinger equation. The same argument could be made, however, regarding the Schrödinger equation~\eqref{eqn:se-grav} applied to neutrons or atoms, where masses are mostly due to the energy of undoubtedly relativistic internal states.
The main difference is not in the relativistic internal energy; but in the diversion from a quantum theory with Galilean spacetime symmetry towards one with the Bargmann group as a symmetry group. This touches on the important query of what constitutes a ``nonrelativistic'' theory, while reinforcing the idea behind the Bargmann superselection rule~\cite{levy-leblondGalileiGroupNonrelativistic1963,wightmanSuperselectionRulesOld1995} that only such dynamical laws with true Galilean symmetry and a constant mass $M$ as Casimir invariant should be considered consistent nonrelativistic theories, whereas superpositions of states of different mass are disallowed and must be treated within a fully Poincaré symmetric framework.

In a similar vein, one may question the whole idea of employing the inconsistency of faster-than-light communication---a relativistic notion---as an argument against a Galilei symmetric model. The contradictions at which one arrives are predominantly due to the \emph{instantaneous} reduction of Alice's state upon Bob's acquisition of which-way information; a concept that, although well established in quantum physics, is in evident conflict with Lorentz covariance, when taken as part of a dynamical law. If one dismisses the implications of faster-than-light signalling on these grounds, one must dismiss most consistency arguments based on the possibility to signal faster than light~\cite{eppleyNecessityQuantizingGravitational1977,gisinStochasticQuantumDynamics1989,bahramiSchrodingerNewtonEquationIts2014,mariExperimentsTestingMacroscopic2016,belenchiaQuantumSuperpositionMassive2018,rydvingGedankenExperimentsCompel2021}, including the arguments against nonlinear Schrödinger dynamics~\cite{gisinStochasticQuantumDynamics1989} and against a fundamentally semiclassical theory~\cite{mollerTheoriesRelativistesGravitation1962,rosenfeldQuantizationFields1963,bahramiSchrodingerNewtonEquationIts2014} for the interaction of gravity with quantum matter, and one is urged to admit the necessity of far-reaching modifications to quantum theory, in order to maintain its consistency.

If, however, there is an expectation of truth behind this type of argument---and without an alternative resolution to our thought experiment---we must make a choice: Either gravity cannot entangle distant masses, for one of a variety of possible reasons, including a fundamentally semiclassical description and a gravity related wave-function collapse. Or the intuitive, yet somewhat naive, nonrelativistic description of internal energy as a dynamical mass does not withstand the scrutiny of mathematical consistency.

\section*{Acknowledgment}
I thank Domenico Giulini, Philip Schwartz, and Marko Toro\v{s} for enlightening discussions. I gratefully acknowledge funding through the Volkswagen Foundation.

\footnotesize\singlespacing

\end{document}